\documentclass[aps,prd,twocolumn,groupedaddress,eqsecnum,amssymb,showpacs,
nofootinbib]{revtex4}
\usepackage{graphicx,mathptm,bm,times,epsfig}

\newcommand{\be}{\begin{equation}}
\newcommand{\ee}{\end{equation}}
\newcommand{\bea}{\begin{eqnarray}}
\newcommand{\eea}{\end{eqnarray}}

\begin{document}

\title{Cosmological Avatars of the Landscape I: Bracketing the SUSY Breaking Scale}

\author{R.~Holman}
\email[]{rh4a@andrew.cmu.edu}
\affiliation{Department of Physics, Carnegie Mellon University, Pittsburgh PA 15213, USA}
\author{ L.~Mersini-Houghton}
\email[]{mersini@physics.unc.edu}
\affiliation{Department of Physics and Astronomy, UNC-Chapel Hill, NC, 27599-3255, USA}
\author{Tomo Takahashi}
\email[]{tomot@cc.saga-u.ac.jp}
\affiliation{Department of Physics, Saga University, Saga 840-8502, Japan}

\date{\today}

\begin{abstract}
We investigate the effects of quantum entanglement between our horizon patch and others due to the tracing out of long wavelength modes in the wavefunction of the Universe as defined on a particular model of the landscape. In this, the first of two papers devoted to this topic, we find that the SUSY breaking scale is bounded both above {\em and} below: $10^{-10} M_{\rm P}\leq M_{\rm SUSY}\leq 10^{-8} M_{\rm P}$ for $GUT$ scale inflation. The lower bound is at least five orders of magnitude larger than the expected value of this parameter and can be tested by LHC physics.
\end{abstract}


\maketitle

\section{Introduction}
\label{sec:intro}

Is the string landscape predictive? This is one of the main points
of contention in theoretical physics today. One
camp~\cite{landscape} claims that the best we will be able to do is
to ask anthropic questions and then hope that these are also the
questions we want to have answered. 

On the other hand, there are those, ourselves included, who believe that it is premature to give up the hope that some dynamical principle might be able to select out at least an interesting subclass of vacua out of the landscape. In particular, asking whether we can find the vacua that allow for consistent inflationary dynamics would be of great interest, especially in light of the WMAP3 data\cite{wmap3}.

In previous work\cite{richlaura1,richlaura2}, we have
advocated the use of the quantum dynamics of gravity to ``thin out''
the herd of vacua, as it were. We showed that the inclusion of the
backreaction of superhorizon matter modes onto the gravitational
degrees of freedom generates a Wheeler-DeWitt master equation from which we could infer that the phase space of stable
inflationary patches is dynamically reduced due to a Jeans instability.
Furthermore, this analysis showed that treating the space of
inflationary initial conditions as if it consisted of regions with
field and gravitational configurations that were in thermodynamic
equilibrium was, in fact, inconsistent \cite{richlaura2} and that perturbative approaches cannot single out our vacua since a perturbed action results only on a rescaling of the vacuum energy $\Lambda$ or some dark radiation contribution\cite{tye,flysoup}. Thus the use of the nonperturbative dynamics of the gravitational and matter degrees of freedom, with the resulting mixed initial state due to nonlocal entanglement, seems a promising avenue as far as the vacuum selection problem is concerned.

To tie this into the landscape, we view the landscape as the natural
configuration space for the wave function of the Universe. The wave
function of the universe propagates on an infinite dimensional minisuperspace which has as its variables the quantities (moduli fields, fluxes of RR field
strengths, etc.) that describe the vacua of the landscape, collectively denoted by $\phi$, together
with the gravitational degrees of freedom \cite{landscape} for the 3-geometries with scale factor $a$ and inhomogenous tensor and scalar fluctuations $\{d_n,f_n\}$.

While the analysis in Refs.~\cite{richlaura1,richlaura2} shows how the initial
conditions for inflation in survivor universes are (super) selected through gravitational quantum dynamics, one can ask whether there are more observationally
distinctive signatures that might arise from this treatment. In fact, we find such signatures and we will elaborate on them in this series of papers. 

In this, the first paper of the series, we show how traces of the combination of gravitational dynamics and quantum entanglement contained in the initial wave-packet describing our universe, together with the requirement that the temperature anisotropies in the CMB were seeded by quantum fluctuations during an inflationary phase allows us to place both upper and {\it lower} bounds on the scale of SUSY breaking. What makes this result particularly exciting is that our lower bound is five orders of magnitude larger than the TeV scale usually assumed; this allows for the possibility that our scenario could be falsified once the LHC starts taking data!

The companion article in this series deals with more detailed imprints that these gravitational effects leave on both the CMB as well as on large scale structure (LSS). 

We first spend some time delineating our model for the landscape, arguing that it probably captures the most important features of the stringy landscape, at least from the point of view of cosmology. Then, in section \ref{sec:backreaction} we exhibit the details of the calculation of backreaction of long
wavelength modes at the onset of inflation and derive the corresponding energy shift of the wavepacket in its phase space trajectory. The energy shift results in a modification to the Friedman equation and the generation of non-gaussian inhomogenities induced by the non-local entanglement left over from primordial times. The physical significance of these results is that our wave-packet preserves unitarity namely, the universe remains in a mixed state even at present, although the scale of the nonlocal entanglement is much larger than the present Hubble radius $r_H$. Observationally, the quantum entanglement between our patch and others and with the superhorizon wavelength modes leaves unique signatures on CMB temperature anisotropy power spectrum as well as on large scale structure (LSS). 

Matching our results to the tight constraints coming from the requirement of flatness of the inflaton potential and the amount of inhomogeneites allowed by CMB data will allow us to bracket the scale of SUSY breaking for our universe. 

\section{A Model of the Stringy Landscape}
\label{sec:model}

The string landscape\cite{landscape} is a vast and complicated space of possible string vacua. They differ in terms of what RR fluxes are turned on, what gauge groups appear in the low energy theory, as well as the values of physical parameters. In particular, there are vacua which preserve supersymmetry as well as those that break it. 

A full analysis of the structure of the landscape is currently beyond our reach. It is expected that there are  {\em at least} $10^{500}$ possible vacua present in the landscape. This suggests a statistical approach to the problem and such a program was begun by Douglas and Denef\cite{douglasdenef}. Their observation was that the matrix of fermion masses in SUSY theories is a complex symmetric matrix and so can be modeled by the so-called CI distribution of Altland and Zirnbauer\cite{altlandzirnbauer}, which leads to a distribution of mass eigenvalues that exhibits level repulsion and hence shows that degenerate eigenvalues are not the generic situation. 

Starting from this observation, one of us (L.~M-H\cite{land1,land2}) constructed a model for the landscape that exhibits it as a lattice of vacua with a distribution of vacuum energies. Different lattices are used in the SUSY and the non-SUSY sectors. In the SUSY sector, the landscape is viewed as a regular lattice and wave functions extend over the whole sector. In the non-SUSY sector, on the other hand, solutions exhibit an Anderson localization around each site as will be discussed further below. It's worth noting that this localization disconnects the two sectors. A wave-function can then be defined that uses this lattice as its configuration space.We found that some statistical aspects of the landscape can be understood via random matrix theory \cite{land1,land2} and that the probability distribution on the landscape phase space with gravity included belongs to the type C universality class \cite{richlaura1} rather than type CI of \cite{altlandzirnbauer}. 

There are good reasons to believe that SUSY is broken in our world. Because of this, the vacuum energy densities can vary widely, being positive, negative or zero. In accord with the distribution of landscape vacua found in Refs.~\cite{douglasdenef,richlaura1}, we take this sector of the landscape to be modeled by a {\em disordered} lattice, where each of the $N$ sites of this lattice is labeled by a mean value $\phi_i$ of the moduli fields, which serves as a collective coordinate for the landscape. In this context, disordering means that the energy density of the vacua have a stochastic distribution which we take to be drawn from the interval  $\left[-W,+W\right]$, where $W\sim M_{\rm Planck}^4$. The disordering of the lattice is enforced by the Gaussian distribution we use to draw energy densities. The width $\Gamma$ of this Gaussian is the disordering strength; we expect that this is related to the amount by which SUSY is broken and we take $M_{\rm SUSY}^8\lesssim \Gamma\lesssim M_{\rm Planck}^8$, where $M_{\rm SUSY}$ is the SUSY breaking scale. We also expect each lattice site to have some structure, corresponding to internal degrees of freedom that capture the distribution of the universality class for the landscape \cite{douglasdenef,richlaura1}; we take this to behave as closely spaced resonances which we label as $\{\phi_i^{n}\}$, where $n$ tags the internal structure. 

Qunatum mechanically we expect tunneling between the vacua to take place. Given this and the disordered nature of the lattice, the interesting aspect of this lattice is that it allows Anderson localization\cite{anderson,land1,land2} to take place around each one of the vacua (lattice sites). For large enough values of the disorder strength $\Gamma$, the majority of the levels are localized so that a semiclassical treatment of their classical trajectories in configuration space is justified. 

If we consider a wave function that has as its configuration space the coordinates of the lattice of non-SUSY vacua, we can use the localization around each site to treat the ensemble of sites as the space of possible initial conditions for the Universe. 

To tie this in to cosmology, we turn gravity on and first consider the minisuperspace determined by the coordinates on the landscape lattice together with the FRW scale factor $a$ so that the wave function of the Universe $\Psi$ is a function of $a,\ \{\phi_i^{n}\}$. The internal degrees of freedom can be used to construct wavepackets in this minisuperspace following the approach of Ref.~\cite{kieferwavepackets}. We take these Anderson localized wavepackets to be Gaussian around each of these vacua with a width $b$ which we expect to be of the order of the supersymmetry breaking scale $M_{\rm SUSY}$ in order to account for the splitting of the zero energy levels. The details of this construction can be found in Refs.~\cite{richlaura1,richlaura2}. 

The minisuperspace described above is not sufficient for our purposes. It has to be expanded to include fluctuations about the various mean values involved. Thus, we allow for metric perturbations about the background FRW geometry as well as perturbations about the scalar degrees of freedom, one of which will be the inflaton; these will be labeled as $\{d_n\},\ \{f_n\}$ respectively. We follow Refs.~\cite{halliwellhawking,kiefer}.

We now have the setup needed to understand the new effects appearing from the landscape. In the next section, we explicitly trace out the long wavelength modes out of the wave function to construct a reduced density matrix $\rho_{\rm red} (a, \phi; a^{\prime}, \phi^{\prime})$ from which we compute the corrections to the energy density that appears in the modification to the Friedmann equation as well as other important quantities.

\section{Calculation of Entanglement and Backreaction Effects}
\label{sec:backreaction}

Our calculation of the entaglement and backreaction contribution to our wavefunction is carried out in the SUSY breaking sector of the landscape, where localization can occur. Each vacuum in this sector thus carries two parameters: the global (Planck) string scale $M_P$ of the theory and the local SUSY breaking scale, $M_{\rm SUSY}$, of the individual vacua under consideration. Both of these scales will appear in the width of the wave-packets localized around each of the vacua in this sector of the landscape and will also show up in the width of the Gaussian suppression for the diagonal and off-diagonal terms of the reduced density matrix. These
describe the entanglement of our inflationary patch with others \cite{richlaura1,richlaura2} and   $M_P,\ M_{SUSY}$ will determine the interference and decoherence lengths of the wavefunction for our patch.

The calculation that follows uses the techniques described in Ref.~\cite{kieferqed} as well as Ref.~\cite{halliwell}, so we summarize below the main results from these works that we will use in the sequel. We are using the higher multipoles both of the scalar field driving inflation as well as (at least in principle) those of the metric to ``measure'' the wavefunction and induce decoherence between different states in a superposition. We treat these multipoles as perturbations on the zeroth order homogenous mode, and only keep their effects to quadratic order in the action. In this we follow Ref.\cite{halliwellhawking,kiefer}.

An intrinsic time $t$ can be defined for WKB wavefunctions using 
$$
\frac{\partial}{\partial t} \equiv \left(\nabla S\right) \cdot \nabla
$$
where $S$ is defined via $\psi_0 \sim C\exp i S$ is the classical action and $\nabla$ is the gradient vector defined on the minisuperspace variables \cite{halliwellhawking,kiefer}.

The reduced density matrix of the system, obtained after tracing out the higher modes, in this approximation, can be written in terms of the density matrix in the absence of fluctuations $\rho_0(a, \phi; a^{\prime}, \phi^{\prime})$ as
\bea
&&\rho(a, \phi; a^{\prime}, \phi^{\prime})= \nonumber \\
&&\rho_0(a, \phi; a^{\prime}, \phi^{\prime}) \prod_{n>0}^N \int df_n \psi_n^* ((a^{\prime}, \phi^{\prime}, f_n) \psi_n(a, \phi, f_n),
\eea
where the wavefunctions $\psi_n$ are solutions of $\hat{H}_n \psi_n = i\dot{\psi}_n$ where $\hat{H}_n$ is given by
\be
\label{eq:hn}
\hat{H}_n= -\frac{\partial^2}{\partial f_n^2} + e^{6 \alpha} \left(m^2+e^{-2 \alpha} (n^2-1)\right) f_n^2
\ee
where $a= \exp \alpha$, $n$ denotes the comoving momentum label in a closed Universe and the $-1$ in the $n^2-1$ term comes from the curvature term. Note that we have not explicitly included here the effects of the tensor metric perturbations on the reduced density matrix, since the procedure is identical to that of tracing out the scalar modes. Besides, the former are simpler in some sense since they are gauge invariant, while the dominant contribution to the spectra and energy corrections comes from the scalar sector (see Ref.~\cite{halliwellhawking} for a more thorough discussion of this point). 

If we make a Gaussian ansatz for $\psi_n$ {\it i.e.}
\be
\psi_n = N(t) \exp(-\frac{1}{2}\Omega_n(t) f_n^2)
\ee
we can insert this into the Schr\"{o}dinger equation for $\psi_n$ to arrive at equations for the normalization factor $N(t)$ and the frequency $\Omega_n(t)$. Taking the results from \cite{kieferqed} we have for the survivor universes \cite{richlaura1}, {\it i.e} in the limit that $m^2\slash H^2\ll 9\slash 4$, where $m$ is the mass of the inflaton field $\phi$:
\be
\Omega_n \simeq \frac{n^2 a^2 \left(n+i a H\right)}{n^2+a^2 H^2} + i \frac{m^2 a^3}{3 H}\equiv \Omega_{n, R} + i \Omega_{n, I}
\ee

The reduced density matrix can be written as \cite{kieferqed}
\be
\rho(\phi+\Delta, \phi)=\rho_0 \exp I 
\ee
with $\Delta =(\phi - \phi^{\prime})$ and 
\be
I=i \left({\rm Tr} \frac{\Omega_I}{\Omega_R}\right)\Delta -\left({\rm Tr}\frac{\left |\Omega^{\prime}\right |^2}{\Omega_R}\right)\frac{\Delta^2}{2}.
\ee
The real part of $I$ corresponds to the cross term of the reduced density matrix which determines the degree of decoherence of our patch, that is, to what extent the phase relations of the entanglement with other patches $\phi^{\prime}$ can be observed.The imaginary part corresponds to the diagonal term of $\rho$ and describes the degree of coherence for the packet in $\phi$ that is the scale at which interference effects become significant. 

From this result we can compute the corrections to the energy, as well as the effects of decoherence between different states present in the density matrix. The decoherence factor is given by \cite{kieferqed}
\be
D = \frac{1}{2} \sum_{n>0} n^2 \frac{\Omega_{n, I}^2}{\Omega_{n, R}^2}.
\ee
This shows up in the reduced density matrix as $\exp \left(-D a^4 (\phi-\phi^{\prime})^2\right)$. 

The entanglement in the initial mixed state induces a shift in the energy of the wavepacket in its trajectory in phase space \footnote{This effect is well known in particle physics: when a charge interacts with a field there is an energy shift in its trajectory.} given by the total hamiltonian which contains the backreaction corrections in our Master equation \cite{richlaura1}, $\mathcal{H}=\mathcal{H_0} + \sum_n \mathcal{H_n}$. The corresponding correction to the energy density of the universe, due to these nonlocal entanglement with other patches and the backreaction of the higher multipoles, originating from this wavepacket, shows as a modification in the Friedman equation and is given by $\Delta E_{\phi} = \mathcal H_{n ,\phi}\slash{V}$ and is given by:
\be
-\Delta E_{\phi} =\frac{1}{V} {\rm Tr}(\Omega_R) + \frac{a H}{V} {\rm Tr}(\frac{\Omega^{\prime}_I}{2 \Omega_R}).
\label{eq:modenergy}
\ee
This energy shift induces a nonlocal correction to the Friedman equation of our universe, given by $V_{eff}= V(\phi) + \Delta E_{\phi}$. We have to unwrap some notation in the above equation. The volume of 3-space is denoted by $V$ (it is just $a^3$). The traces are sums over the modes, $\Omega_{R, I} = \Omega_{n, R, I}$ and the prime denotes a derivative with respect to the scale factor $a$.

The interference or coherence length can also be calculated and it is given by $l_{\rm coh}^2 = A_1$, where
\be
H A_1 = {\rm Tr}(\frac{\Omega_I}{\Omega_R}).
\ee

To compute the relevant traces, we approximate the sums over mode numbers by integrals. The only question is what to take as the UV and IR cutoffs on these integrals. Since we are tracing out superhorizon modes, the {\it upper} limit on $n$ should be $a H$. 

The lower limit is a somewhat more subtle question to deal with since it is related to the well known IR divergence problem in quantum gravity. We would argue here that a physically well motivated choice for the IR cuttoff is given by the SUSY breaking scale  $b$ which determines the vacuum energy. The main point is that we are constructing wave-packets localized on a landscape vacuum whose width is determined by scale $b$. If we probe this wavepacket on scales shorter than $b$ this would destroy its quantum coherence and localization by exciting the system in such a manner that the wave packet would spread over many vacua and disintegrate into its many components. Another way to say this is that on scales larger than its width, the wavepacket can be approximated by a classical particle so that a classical trajectory in phase space and decoherence for our universe are assured. But if the characteristic scale of disturbance is shorter than the characteristic size of the system given by its width $b$ then the interference effects among its many components become significant, a proccess which destroys decoherence. For this reason we take $k=ab$ as the IR cutoff for the scale of entanglement of our inflaton patch. Doing this gives us:
\be
H A_1 = i a \int_{a b}^{a H} dn \ \frac{m^2(n^2+a^2 H^2)+ 3 n^2 H^2}{3 n^3 H}.
\ee
The integral yields:
\be
H A_1 =  -i a \left\{\left(\frac{m^2}{3 H}+H\right)\ln\frac{b}{H}-\frac{m^2 H}{6}\left(\frac{1}{b^2}-\frac{1}{H^2}\right)]\right\}.
\ee
We also find
\bea
D &\sim& a^2  \int_{a b}^{a H} dn\ n\ \left( \frac{m^2(n^2+a^2 H^2)+ 3 n^2 H^2}{3 n^3 H} \right)^2\nonumber\\
&=& -a^2 H^2 \left[\left(1+\frac{m^2}{3 H^2}\right)^2 \ln\frac{b}{H}-\frac{a^4}{36}\left(\frac{1}{b^4}-\frac{1}{H^4}\right)\right . \nonumber \\
&&\left .-\frac{1}{3}\left(1+\frac{m^2}{3 H^2}\right)\left(\frac{1}{b^2}-\frac{1}{H^2}\right)\right].
\eea

Finally we compute $\Delta E_{\phi}$. This requires us to compute two traces: $A_3 \equiv{\rm Tr}(\Omega_R)$ and 
 $A_4\equiv {\rm Tr}(\Omega^{\prime}_I\slash 2 \Omega_R)$. We do these in turn.
\be
A_3 = \int_{a b}^{a H} dn\ n \frac{n^2 a^2} {n^2+a^2 H^2} = -\frac{a^4 H^2}{2} \left[\left(\frac{b^2}{H^2}-1\right)-\ln\frac{b}{H}\right].
\ee

A similar calculation yields 
\be
A_4 = \frac{3 H}{2} A_1 +\frac{a H^2}{2} \ln\left[\frac{2 b^2}{b^2 + H^2}\right].
\ee

\section{Landscape Constraints on the SUSY Breaking Scale}
\label{sec:susybreaking}

The fact that backreaction effects can help solve the problem of
inflationary initial conditions is fascinating in and of itself.
However, this will be academic unless we can argue that
there are falsifiable consequences arising from these effects. In this
section, we show that our knowledge of the CMB power spectrum can
bound some of the parameters which are a part of our description of the  landscape. One such parameter is $b$. As described above, it describes the width of the wavepackets constructed around each vacuum in the landscape using the internal excitations around each vacuum and is related to the SUSY breaking scale. What we will find is that $b$ can be related to the value of the quadrupole of CMB \cite{smoot}. On the other hand, it is also related to the amount of quantum interference, as determined by the reduced density matrix, between our horizon patch and others.

For definiteness we will follow Ref.~\cite{laurakatie} and use the following inflaton
potential 
\be
 \label{eq:infpot} V(\phi) = V_0 \exp\left(-\lambda
\frac{\phi}{M_{\rm P}}\right). \label{eq:V_inf} 
\ee 
This inflationary potential can arise in SUGRA models, which is the reason we chose this particular example for illustration. However, our results will be valid for generic potentials that allow for an inflationary phase.

When the backreaction effects are included, due to the energy shift in the WKB trajectory of our wave-packet in phase space the relevant Friedmann equation becomes modified as follows
\be
\label{eq:modfried} H^2 = \frac{1}{3 M_{\rm P}^2}
\left[V(\phi)+\frac{1}{2} \left(\frac{V(\phi)}{3 M_{\rm
P}^2}\right)^2 F(b,V)\right]\equiv \frac{V_{\rm eff}}{3 M_{\rm P}^2}
\ee
where
\bea \label{eq:corrfactor}
F(b,V) &=& \frac{3}{2} \left(2+\frac{m^2M^2_{\rm P}}{V}\right)\log \left( \frac{b^2 M_{\rm P}^2}{V}\right)\nonumber \\
&-&\frac{1}{2} \left(1+\frac{m^2}{b^2}\right) \exp\left(-3\frac{b^2
M_{\rm P}^2}{V}\right). \eea

Note that here, we have approximated the dependence on the Hubble parameter $H$ contained in Eq.~(\ref{eq:modenergy}) by the inflaton potential $H^2 \approx V(\phi)$.
We have taken $8\pi G_N = M_{\rm P}^{-2}$, $b=M_{SUSY}$ is the SUSY breaking scale and $m^2=V^{\prime\prime}(\phi)$. There are two types of corrections appearing in our modification to the Friedmann equation. The term involving the exponential arises from the nonlocal entanglement of our horizon patch with others. While the calculation in Sec.~\ref{sec:backreaction} only obtains the first two terms of an expansion of the exponential in terms of $b^2\slash H^2$, we exponentiated it, anticipating that it corresponds to a tunneling type of correction. This is based on the fact that for our ansatz of the wavefunction we can formally write the energy corrections to be of the familiar form coming from particle creation: $\mathcal{H_{\phi}} \approx Tr(\omega + 2\omega|\beta|^2 )$ with $\Omega \approx \omega + i\dot{\omega}\slash 2\omega$ and $\beta^2$ the exponential term here, which formally corresponds to particle creation \cite{kieferqed}. Since $b^2\slash H^2$ will be small as shown in \cite{richlaura2}, doing this will not cause any inaccuracies. 

The term involving the logarithm incorporates the effects of superhorizon massive fluctutations. Note that both of these corrections involve  a nontrivial time dependent function of the coupling between the effects due to the inflaton potential $V(\phi)$ and superhorizon fluctuations represented by $b$. This will give rise to subtle effects on large scale structure, as we discuss in the second paper in this series \cite{avatars2}.

The primordial power spectrum is given by \be P_\mathcal{R} = \frac{1}{75 \pi^2 M_{\rm
P}^2 } \frac{V_{\rm eff}^3}{V_{\rm eff}^{'2}}. \ee For the potential given by Eq.~(\ref{eq:infpot}), we have
\be P_\mathcal{R}^0 = \frac{1}{75\pi^2  M_{\rm P}^2
} \frac{V_0}{ \lambda^2 M_{\rm P}^4} \ee The scalar spectral index
is given by $n_s -1 =  -\lambda^2$. Modifications in the Friedmann equation result in a running of the spectral index $n_s = n_s^{0} +\delta n_s$, as we describe below.

In our case, i.e., with the effective potential $V_{\rm eff}$, the
situation becomes more complicated. Now the solution for the inflaton field 
becomes 
\bea \phi &=& \lambda M_{\rm pl} \left[ 1 +
\frac{1}{2} \frac{1}{ 3M_{\rm pl}^2} \left(\frac{V_0}{3 M_{\rm
pl}^2} \right)
\right. \nonumber \\
&& \left. \times \left\{ 3 \left( 2 + \frac{m^2 M_{\rm pl}^2}{V_0}
\right) \log \left( b \sqrt{ \frac{3M_{\rm pl} }{V_0}} \right)
\right.  \right. \nonumber  \\
&& \left. \left. -\frac{1}{2} \left( 1 + \frac{m^2}{b^2} \right)
e^{- 3M_{\rm pl}^2 b^2 /V_0}  \right\} \right]^{-1}
\log \left( \frac{k}{k_{\rm ref}} \right)  \nonumber \\
\label{eq:fieldsol}
\eea 
where $k_{\rm ref} \simeq (4000 ~{\rm Mpc})^{-1}$.

Define $3M_p^{2}/F(b,V) \equiv \sigma(b,\phi)$ and denote the energy correction $V^2 / \sigma = f(b,V)$. The modified Friedmann equation can then be written as
\be
\label{eq:friedshorthand}
3M_p^{2} H^2 = V + f(b,V)
\ee
Notice that $f(b,V)$ is a negative function, so that we are restricted to the regime for which the right hand side of this equation is positive.

We are now ready to derive the cosmological bounds for the local SUSY breaking scale in our patch.

\subsection{Flatness of the Inflaton Potential}:
 
As is well known, for a succesful stage of inflation to occur, inflationary potentials have to be 'fine-tuned' such that they satisfy the flatness condition \cite{adamskatie}
\be
\Delta V \slash (\Delta\phi)^4 \le O(10^{-7})
\ee

For GUT scale inflation with $\Delta\phi\simeq {\cal O}(M_P)$, the expontential potential type considered here would satisfy this condition, if we choose parameters such that for example: $V_0 \simeq 10^{-9} M_P^{4}, \ \lambda \simeq 0.1$.

When the primordial effects of entanglement and backreaction of superhorizon matter perturbations on the inflaton potential are taken into account it is the effective potential  $V_{eff} = V + V^2 \slash \sigma(b,V)$ that must satisfy this condition. Now note that we can approximate $\sigma(b,V)= M_P^4 /(\Delta N_b -m^2 \slash b^2)$, with $\ln(3M_P^{2} \ b^2 \slash V)\simeq 2 \ln(k_b \slash k)\simeq \Delta N_{b}$, where $\Delta N_{b}$ is the number of e-folds before the end of inflation at which the scale $b$ leaves the horizon and we take $e^{-3M_p^2 b^2 \slash V} \simeq 1$. Using this, the flatness condition gives $\sigma > 10^{-9} M_P^{4}$ that places a lower bound on the $SUSY$ breaking scale
\be
b \ge 10^{-10} M_P.
\label{eq:lowerbound}
\ee

\subsection{Constraints from CMB experiments:} 
The second condition on the SUSY breaking scale comes from the TT power spectrum of the CMB. Corrections disturb slow roll inflation. Inhomogeneities on scales larger than the horizon induce gradients and shear across our horizon \cite{smoot} that affect the Newtonian background potential. The contribution from these corrections to the quadropole is constrained to be \cite{smoot,wmap3}
\bea
\label{eq:quadrupole}
&(\nabla T / T)_{\rm quad}&\approx r_H^{2}\nabla^2 \delta\phi \nonumber\\
&=(c k_{1} / H_0 )^2\delta\phi&\approx 0.5 (r_H/L_{1})^2 (\delta\rho / \rho)_{1}.
\eea

The subscript $1$ denotes the wavenumber/scale where inhomogeneities arising from the entanglement $\sigma$ dominate, $0$ denotes present day values and $\delta\phi$ is the $\sigma$-induced quadrupole Newtonian potential. Notice that the quadrupole contribution from entaglement to the amplitude of anisotropies turns out to be the interference length of the wavepacket obtained from the width of the diagonal terms of the density matrix, as derived in the previous section:  $(A_1/aH) = (L_1\slash r_H )^{-2}$. 

Taking into account the possibility that other fluctuations from such as 
curvaton \cite{curvaton} and modulated reheating \cite{inhomdecay} can contribute to the primordial 
fluctuation, we take 
$\delta\rho / \rho \lesssim 10^{-5}$ then the bound from Eq.~(\ref{eq:quadrupole}) reads: $(L_1 \slash H_0)^2 =(A_1\slash aH) >10^5$ thus $b^2/6m^2 \le 10^{-5}$ or 
\be
b \le 10^{-8} M_P 
\label{eq:upperbound}
\ee 
We have thus derived cosmological upper and lower bounds on the SUSY breaking scale $b=M_{\rm SUSY}$
\be
10^{-10} M_P < b < 10^{-8} M_P
\label{eq:susybounds}
\ee

The cosmological bounds derived here are relevant for the GUT scale inflation. 
However, as it can be seen from Eqs.~(\ref{eq:lowerbound}) and (\ref{eq:upperbound}) it is straightforward to derive the lower and upper bound for the SUSY breaking scale as a function of $V$ by our approach here for any scale of inflation, namely:
\be
\frac{V}{ M_p^{4}} < \frac{b^2}{ m^2} <  10^{-5}
\label{eq:general}
\ee
where $V, m^2$ are the inflaton potential and mass squared respectively at any scale.The scale of inflation is bound by the reheating temperature to be at least of order TeV. In this case the lowest possible bound on the SUSY breaking scalebecomes $b > 10{-15} M_p$, a value which might be within the reach of LHC. Henceif $b$ is observed at LHC scales, we would learn valuable information not only about SUSY and Higgs physics but also information about the inflation scale itself!

Two comments are in order. First it is very interesting that the induced quadrupole anisotropy scale is given by the intereference length of the wavepacket. This provides a natural physical explanation behind the channel of the induced inhomogeneities on the LSS. It also sheds light on the reason why $b$ is the IR cutoff for our patch.  The underlying significance of this result is that our inflaton bubble is a classical world roughly up to the scales given by the interference length $L_{1} > 10^{2.5} H_0$. Beyond  this scale, we can expect to see strong quantum interference effects associated with the nonlocal entangelement in the mixed state of our universe and with the fundamentally quantum nature of the fluctuations.

\begin{figure}[!htbp]
\begin{center}
\raggedleft \centerline{\epsfxsize=3.5in
\epsfbox{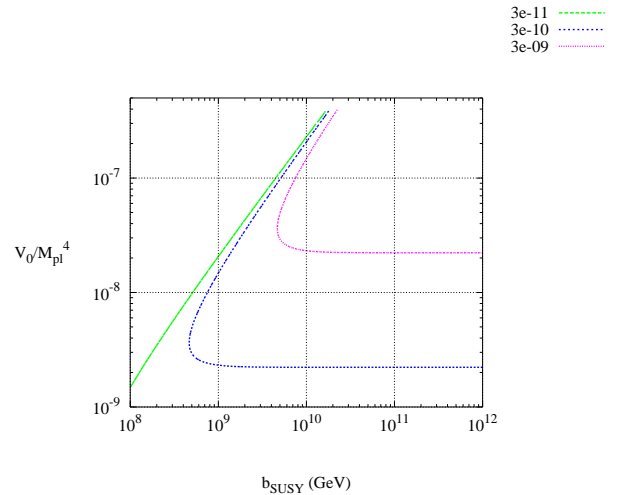}} 
\caption{Contours of $P_k$ at $k=0.002
~{\rm Mpc}$ are shown. The value of $\lambda$ is fixed as $\lambda =
0.1$ in this figure. } \label{fig:fig1}
\end{center}
\end{figure}

Perhaps more surprising, the cosmological bounds obtained above place tight constraints on the SUSY breaking scale; they appear to force it to be about 5 orders of magnitutde larger than the normal TeV expectation. These bounds soon will be complemented by ones from the LHC. If our approach is correct and that it is indeed true that cosmology requires that SUSY be broken at very high energy scales then scenarios such as Split Supersymmetry \cite{nimasplit} may be the only way to make use of supersymmetry to deal with the hierarchy problem. 
We checked these bounds by performing a numerical analysis which confirmed are analytical findings and the fact that the cosmological bounds found here are very stringent indeed!
The results of the numerical analysis are shown in Fig.\ref{fig:fig1} where we have ploted the contours of SUSY breaking versus GUT scale inflaton potential $V$.

The fact that cosmology rather than particle physics can place such tight bounds on the SUSY breaking scale and require it to be so much higher than expected coupled with the fact the LHC will soon be able to test our SUSY breaking bounds directly is an exciting possibility indeed!

\section{Conclusions}
\label{sec:conclusions}

No scientific theory can be considered to be on firm grounds unless it makes predictions that can be tested. What can be said about the quantum gravity string landscape as a candidate for the underlying theory for the early universe? We took up these issues in a series of papers \cite{richlaura1,richlaura2,land1,land2} by proposing that the landscape provides the phase space for the ensemble of initial patches known as the multiverse. We then allowed the wavefunction of the universe to propagate through this structure  in order to find out which one of the vacua would be selected as our initial patch, {\it i.e.}  to  address the issue of the selection of the initial conditions from the point of view of a superselection rule emerging from the quantum dynamics of gravity \cite{richlaura1}.

Our picture of the landscape and how the wavefunction of the Universe is affected by quantum gravity effects such as  nonlocal entanglement with other patches gives rise to observational consequences, shown here and its companion paper \cite{avatars2}, that may in fact explain some of the strange features found in studies of the CMB as well as LSS. 
There is no way one could phenomenologically guess the nonlocal entanglement between $b$ and $V(\phi)$ in the highly nontrivial corrections to the Friedman equation, Eqs.~(\ref{eq:corrfactor}) and (\ref{eq:modfried}), and its subsequent unique signatures on CMB and LSS, derived in Sec.III. Our proposal has thus not just provided a good working model for derving the quantum gravity effects, left from the early times, in the multiverse phase space of the landscape, thereby leading the way for {\it a dynamic rather than anthropic approach} to the selection of our universe. But, as we showed here and in the next   paper in this series \cite{avatars2}, it also makes predictions that can be tested by observations. By probing into the underlying structure of the initial state and confronting the issues of the origin of our universe from the physics of quantum gravity, we have shed some light into the interrelation between the SUSY breaking scale and the size of nonlocality of quantum entanglement.

To summarize our results, we have seen that the requirement of having a sufficiently flat inflationary potential {\it after} the modifications to the Friedmann equation are taken into account, coupled with the known value of the COBE quadrupole put stringent bounds on the energy scale related to the structure of vacua in the non-SUSY part of the landscape. In our picture, this is the actual SUSY scale, and we find that it has to be significantly larger (five to eight orders of magnitude larger) than studies of the hierarchy problem would have required. The LHC will soon be able to test this statement, once again showing the tight interconnectivity between particle physics and cosmology.
It is amazing how a coherent cosmological picture of the early universe can predict such tight bounds on a particle physics parameter by relating it to imprints of the nonlocal entanglement of our universe with other horizon patches on astrophysical observables. More interestingly, we make predictions which are within the observational limits of current experiments such SDSS, WMAP, HST, as well as within those of the upcoming Planck, LISA and lensing experiments  and will soon be compared againts the LHC results.

In the second paper in this series \cite{avatars2} we will show how the effects of entanglement and backreaction due to the superhorizon modes has very interesting, and more importantly, testable effects on the CMB and large scale structure.  

Our feeling in this enterprise is that our model of the landscape contains enough of the coarse features of the true stringy landscape to be a reliable model of it. What we find rather startling is the plethora of cosmological manifestations of the physics of the landscape and how amenable to observation they are. We expect that even if our model does not yet capture the finer details of the landscape, the strategy of using the landscape as the space of initial states for the wavefunction of the universe and following its evolution in the manner described both here and in our previous work promises to be a fruitful approach.

\begin{acknowledgments}
R.~H. was supported in part by DOE grant DE-FG03-91-ER40682. He would also like to thank the Perimeter Institute for their generous hospitality while this work was in progress. L. ~M-H was supported in part by DOE grant DE-FG02-06ER1418 and NSF grant PHY-0553312. 
\end{acknowledgments}

\end{document}